\documentclass[12pt]{article}
\usepackage{setspace}
\usepackage{epsfig}
\usepackage{amsmath}
\usepackage{amssymb}
\usepackage{graphicx}
\def\be{\begin{equation}}
\def\ee{\end{equation}}
\def\ba{\begin{array}}
\def\ea{\end{array}}
\def\beqn{\begin{eqnarray}}
\def\eeqn{\end{eqnarray}}

\def\bt{\begin{tabular}}
\def\et{\end{tabular}}
\def\bc{\begin{center}}
\def\ec{\end{center}}

\begin{document}

\title{General lepton textures and their implications}

\author{Gulsheen Ahuja$^*$, Samandeep Sharma$^@$, Priyanka Fakay, Manmohan
Gupta\\ {\it Department of Physics, Panjab University,
 Chandigarh, India.}\\
{\it $^@$Department of Physics, GGDSD College, Chandigarh,
India.}\\
\\{\it $^*$gulsheen@pu.ac.in}}

\maketitle

\begin{abstract}
The present work attempts to provide an overview of  texture
specific lepton mass matrices. In particular, we summarize the
findings of some recent analyses carried out within non flavor
basis, wherein a parallel texture structure for the lepton and
neutrino mass matrices is considered.
\end{abstract}

\section{Introduction}
In the last decade, we have almost reached `precision' level for
the measurement of neutrino oscillation parameters, including the
recently measured mixing angle $\theta_{13}$. This has led to the
need of a more intense activity towards understanding the pattern
of neutrino masses and mixings which is quite different from the
corresponding quark mixing case. In the absence of a theory
providing a viable understanding of these issues, most of the
phenomenological work is carried out within the general premises
of `bottom-up' approach. As an example of this approach, texture
specific lepton mass matrices have been tried with a good deal of
success. In particular, several attempts
\cite{flbaatt1}-\cite{flbaatt8} have been made to understand the
neutrino mixing data by formulating the phenomenological mass
matrices with charged lepton matrix being diagonal, usually
referred to as the flavor basis case. In addition, for both
Majorana as well as Dirac neutrinos, some attempts
\cite{minireview}-\cite{singreview} have also been made to explain
the neutrino mixing data by considering texture specific
structures for both the charged lepton and the neutrino mass
matrices, referred to as the non flavor basis case. It may be
noted that the non flavor basis enables quarks and leptons to be
treated at the same footing and also to explore the possibility to
arrive at a minimal set of fermion mass matrices which are
compatible with the latest mixing data.

It is now well known, that, in the leptonic sector, the search for
viable mass matrices is complicated by the `smallness' of neutrino
masses. The most popular explanation for this smallness is the
see-saw mechanism \cite{seesaw}-\cite{seesaw5} which requires the
neutrinos to be Majorana fermions. However, at present, neither
the Majorana nature is established nor can we rule out the Dirac
nature of neutrinos. On theoretical grounds, the existence of
small Dirac masses requires the corresponding Yukawa couplings to
be exceptionally small compared to their charged counterparts. The
Dirac neutrino mass, although seemingly `unnatural', can be
explained by additional $U(1)_{B-L}$ symmetry of the Lagrangian
which forbids Majorana mass term for the neutrinos. Apart from SM,
this possibility can be realized in many of the models
\cite{singreview} such as supersymmetry, superstring, supergravity
and large extra dimensions. Keeping in mind that Dirac neutrinos
are still not ruled out, in the present work we discuss texture
specific mass matrices for both Dirac as well as Majorana
neutrinos.

In particular, we have presented an overview of some recent
analyses \cite{minireview, arxive1,arxive2} wherein texture
specific lepton mass matrices have been considered in the non
flavor basis for Dirac as well as Majorana neutrinos. In the
following section, we first present the relation between lepton
mass matrices and mixing matrix. The present experimental status
of the neutrino mixing parameters have been give in Section 3. A
brief summary of texture 6, 5 and 4 zero lepton mass matrices has
been presented in Section 4. Finally, Section 5 summarizes our
conclusions.

\section{Lepton mass matrices and PMNS matrix}
For the case of neutrinos, it is important to note that these may
have either the Dirac masses or the more general Dirac-Majorana
masses. A Dirac mass term can be generated by the Higgs mechanism
with the standard Higgs doublet. In this case, the neutrino mass
term can be written as
\begin{equation}
 \overline{\nu}_{a_{L}} M_{\nu D} {\nu}_{a_{R}} + h.c.,
\end{equation}
where $a$ = $e$, $\mu$, $\tau$. $\nu_e$, $\nu_\mu$, $\nu_\tau$ are
the flavor eigenstates and $M_{\nu D}$ is a complex $3\times 3$
Dirac mass matrix. As mentioned earlier, in the non flavor basis,
both the charged lepton and the neutrino mass matrices are
considered having the same texture structure \cite{minireview},
e.g.,
\be
 M_{l}=\left( \ba{ccc}
0 & A _{l} & 0      \\ A_{l}^{*} & D_{l} &  B_{l}     \\
 0 &     B_{l}^{*}  &  C_{l} \ea \right), \qquad
M_{\nu D}=\left( \ba{ccc} 0 &A _{\nu} & 0      \\ A_{\nu}^{*} &
D_{\nu} &  B_{\nu}     \\
 0 &     B_{\nu}^{*}  &  C_{\nu} \ea \right),
 \label{frzmm5}
 \ee
$M_{l}$ and $M_{\nu D}$ respectively corresponding to hermitian
Dirac-like charged lepton and neutrino mass matrices. It may be
noted that each of the above matrix is texture 2 zero type with
$A_{l(\nu)} =|A_{l(\nu)}|e^{i\alpha_{l(\nu)}}$
 and $B_{l(\nu)} = |B_{l(\nu)}|e^{i\beta_{l(\nu)}}$, in case these
 are symmetric then $A_{l(\nu)}^*$ and $B_{l(\nu)}^*$ should be
 replaced by $A_{l(\nu)}$ and $B_{l(\nu)}$, as well as
 $C_{l(\nu)}$ and $D_{l(\nu)}$ should respectively be defined as $C_{l(\nu)}
 =|C_{l(\nu)}|e^{i\gamma_{l(\nu)}}$ and $D_{l(\nu)}
 =|D_{l(\nu)}|e^{i\omega_{l(\nu)}}$.

The texture 6 zero mass matrices can be obtained from the above
mentioned matrices by taking both $D_l$ and $D_{\nu}$ to be zero,
which reduces the matrices $M_{l}$ and $M_{\nu D}$ each to texture
3 zero type. Texture 5 zero matrices can be obtained by taking
either $D_l=0$ and $D_{\nu}\neq 0$ or $D_{\nu}=0$ and $D_l \neq
0$, thereby, giving rise to two possible cases of texture 5 zero
matrices, referred to as texture 5 zero $D_l=0$ case pertaining to
$M_l$ texture 3 zero type and $M_{\nu D}$ texture 2 zero type and
texture 5 zero $D_{\nu}=0$ case pertaining to $M_l$ texture 2 zero
type and $M_{\nu D}$ texture 3 zero type.

It should be noted that in the case of texture 6 zero and texture
4 zero mass matrices we can have parallel structures for both the
neutrino mass matrix and the charged lepton mass matrix, however,
for the case of texture 5 zero mass matrices, one cannot have
parallel structures. To consider all possible textures, we have
considered only those possibilities which are compatible with the
`Weak Basis' transformations \cite{branco,costa}. To this end, in
Table 1, we have presented all possible texture 2 zero mass
matrices, from which we can derive texture 6 zero, 5 zero and 4
zero mass matrices for the discussion.

\begin{table}[hbt]
{\renewcommand{\arraystretch}{1.0} \setlength\arraycolsep{0pt}
\bt{|c|c|c|c|c|} \hline
 & Class I  & Class II  & Class III & Class IV  \\ \hline &&&&\\
 a & $\left ( \ba{ccc} {\bf 0} & Ae^{i\alpha} &
{\bf 0} \\ Ae^{-i\alpha}  & D & Be^{i\beta} \\ {\bf 0} &
Be^{-i\beta}  & C \ea \right )$  &
 $\left ( \ba{ccc} D & Ae^{i\alpha} &
{\bf 0} \\ Ae^{-i\alpha}  & {\bf 0} &  Be^{i\beta} \\ {\bf 0} &
Be^{-i\beta}  & C \ea \right )$     &
 $\left ( \ba{ccc} {\bf 0} & Ae^{i\alpha} &
 De^{i\gamma}\\ Ae^{-i\alpha}  &{\bf 0}  & Be^{i\beta} \\ De^{-i\gamma} & Be^{-i\beta}  &
C \ea \right )$ & $\left ( \ba{ccc} A & {\bf 0} & {\bf 0} \\ {\bf
0}  & D & Be^{i\beta}\\ {\bf 0} & Be^{-i\beta} & C \ea \right )$\\
& & & &\\

b &  $\left ( \ba{ccc} {\bf 0} & {\bf 0}  & Ae^{i\alpha} \\ {\bf
0}  & C & Be^{i\beta} \\  Ae^{-i\alpha} & Be^{-i\beta}  & D \ea
\right )$  &
 $\left ( \ba{ccc} D & {\bf 0} & Ae^{i\alpha}
 \\ {\bf 0} & C & Be^{i\beta} \\  Ae^{-i\alpha} & Be^{-i\beta}  &
{\bf 0} \ea \right )$     & $\left ( \ba{ccc} {\bf 0} &
De^{i\gamma} & Ae^{i\alpha}
\\  De^{-i\gamma}  & C  & Be^{i\beta} \\  Ae^{-i\alpha} & Be^{-i\beta}  &
 {\bf 0}\ea \right )$ & $\left (\ba{ccc}C & {\bf 0} & Be^{i\alpha}
 \\ {\bf 0}  & A & {\bf 0} \\ Be^{-i\alpha}  & {\bf 0}  &
D \ea \right )$\\   & & && \\

c &  $\left ( \ba{ccc} D & Ae^{i\alpha} &
  Be^{i\beta}\\ Ae^{-i\alpha}  & {\bf 0}  & {\bf 0} \\  Be^{-i\beta} & {\bf 0}
& C \ea \right )$  & $\left ( \ba{ccc} {\bf 0} & Ae^{i\alpha} &
 Be^{i\beta} \\ Ae^{-i\alpha}  & D & {\bf 0}  \\  Be^{-i\beta}  & {\bf 0} &
C \ea \right )$     & $\left ( \ba{ccc} {\bf 0} & Ae^{i\alpha} &
Be^{i\beta}
\\ Ae^{-i\alpha} & {\bf 0} & De^{i\gamma} \\ Be^{-i\beta}  & De^{-i\gamma} &
C \ea \right )$ & $\left ( \ba{ccc} C & Be^{i\alpha} & {\bf 0} \\
Be^{-i\alpha}  & D & {\bf 0} \\ {\bf 0} & {\bf 0}  & A \ea \right
)$ \\  & & && \\

d &  $\left ( \ba{ccc} C & Be^{i\beta} & {\bf 0}
 \\ Be^{-i\beta} & D & Ae^{i\alpha}  \\ {\bf 0} & Ae^{-i\alpha} & {\bf 0}
 \ea \right )$   &  $\left ( \ba{ccc} C & Be^{i\alpha} &
{\bf 0} \\ Be^{-i\alpha}  & {\bf 0} &  Ae^{i\beta} \\ {\bf 0} &
Ae^{-i\beta}  & D \ea \right )$   &  $\left ( \ba{ccc} {\bf 0} &
Be^{i\alpha} &
 Ce^{i\gamma}\\ Be^{-i\alpha}  &{\bf 0}  & Ae^{i\beta} \\ Ce^{-i\gamma} & Ae^{-i\beta}  &
D \ea \right )$   & $\left ( \ba{ccc} A & {\bf 0} & {\bf 0} \\
{\bf 0}  & C & Be^{i\beta}\\ {\bf 0} & Be^{-i\beta} & D \ea \right
)$ \\  & & && \\

e &  $\left ( \ba{ccc} D & Be^{i\beta} & Ae^{i\alpha}
  \\ Be^{-i\beta}  & C & {\bf 0} \\ Ae^{-i\alpha} & {\bf 0}
& {\bf 0} \ea \right )$  &  $\left ( \ba{ccc} C & {\bf 0} &
Be^{i\alpha}
 \\ {\bf 0} & D & Ae^{i\beta} \\  Be^{-i\alpha} & Ae^{-i\beta}  &
{\bf 0} \ea \right )$   & $\left ( \ba{ccc} {\bf 0} &
Ce^{i\gamma} & Be^{i\alpha}
\\  Ce^{-i\gamma}  & D & Ae^{i\beta} \\  Be^{-i\alpha} & Ae^{-i\beta}  &
 {\bf 0}\ea \right )$ & $\left (\ba{ccc}D & {\bf 0} & Be^{i\alpha}
 \\ {\bf 0}  & A & {\bf 0} \\ Be^{-i\alpha}  & {\bf 0}  &
C \ea \right )$ \\  & & && \\

f &  $\left ( \ba{ccc} C & {\bf 0} & Be^{i\beta}
 \\ {\bf 0} & {\bf 0} & Ae^{i\alpha}  \\ Be^{-i\beta} & Ae^{-i\alpha} & D
 \ea \right )$  & $\left ( \ba{ccc} {\bf 0} & Be^{i\alpha} &
 Ae^{i\beta} \\ Be^{-i\alpha}  & C & {\bf 0}  \\  Ae^{-i\beta}  & {\bf 0} &
D \ea \right )$  & $\left ( \ba{ccc} {\bf 0} & Be^{i\alpha} &
Ae^{i\beta}
\\ Be^{-i\alpha} & {\bf 0} & Ce^{i\gamma} \\ Ae^{-i\beta}  & Ce^{-i\gamma} &
D \ea \right )$ & $\left ( \ba{ccc} C & Be^{i\alpha} & {\bf 0} \\
Be^{-i\alpha}  & D & {\bf 0} \\ {\bf 0} & {\bf 0}  & A \ea \right
)$\\
  \hline
\et \vspace{0.7 cm}} \caption{Table showing various `Weak Basis'
transformation compatible texture 2 zero possibilities categorized
into four distinct classes and their permutations given by
a,b,c,d,e,f. } \label{52z}
\end{table}

Coming to the diagonalization of lepton mass matrices, similar to
the quark sector, these can also be diagonalized by bi-unitary
transformations, e.g.,
\be
M^{diag}_{\nu D}= U^{\dagger}_{\nu L} M_{\nu D} U_{\nu R}  = {\rm
Diag} ( m_1, ~ m_2, ~ m_3 ), \label{dirtr} \ee where $U_{\nu L}$
and $U_{\nu R}$ are unitary matrices and $M^{diag}_{\nu D}$ is a
diagonal matrix. The corresponding mixing matrix obtained, known
as Pontecorvo-Maki-Nakagawa-Sakata (PMNS) or lepton mixing matrix
 $ V_{PMNS} $, is given as \be V_{\rm PMNS}=
V^{\dagger}_{l_{L}} V_{\nu_{L}}, \ee where $V^{\dagger}_{l_{L}}$
and $V_{\nu_{L}}$ correspond to the diagonalization
transformations of lepton and  neutrino mass matrices
respectively. The $V_{PMNS}$  expresses the relationship between
the neutrino mass eigenstates and the flavor eigenstates, e.g.,
  \be
\left( \ba{c} \nu_e
\\ \nu_{\mu}
\\ \nu_{\tau} \ea \right)
  = \left( \ba{ccc} V_{e1} & V_{e2} & V_{e3} \\ V_{\mu 1} & V_{\mu 2} &
  V_{\mu 3} \\ V_{\tau 1} & V_{\tau 2} & V_{\tau 3} \ea \right)
 \left( \ba {c} \nu_1\\ \nu_2 \\ \nu_3 \ea \right),  \label{nm1}  \ee
where $ \nu_{e}$, $ \nu_{\mu}$, $\nu_{\tau}$ are the flavor
eigenstates ; $ \nu_1$, $ \nu_2$, $ \nu_3$ are the mass
eigenstates and the $3 \times 3$ mixing matrix is leptonic mixing
matrix. For the case of three Dirac neutrinos, in the Particle
Data Group (PDG) parameterization, involving three angles
$\theta_{12}$, $\theta_{23}$, $\theta_{13}$ and the Dirac-like CP
violating phase ${\delta}_l$ the mixing matrix has the form
\begin{eqnarray}
V_{\rm PMNS}=   \left (
  \begin{array}{ccc}
    c_{12} c_{13} & s_{12} c_{13} & s_{13} e^{-i {\delta}_l} \\
    -s_{12} c_{23} - c_{12} s_{23} s_{13} e^{i {\delta}_l} & c_{12} c_{23} - s_{12}
    s_{23} s_{13}e^{i {\delta}_l} & s_{23} c_{13} \\
    s_{12} s_{23} - c_{12} c_{23} s_{13}e^{i {\delta}_l} & -c_{12} s_{23} - s_{12}
    c_{23} s_{13}e^{i {\delta}_l} & c_{23} c_{13}
  \end{array}
  \right ),
\label{ch1pmns2}
\end{eqnarray}
with $s_{ij} = {\rm sin}\theta_{ij}$, $c_{ij} = {\rm
cos}\theta_{ij}$.

The neutrino might be a Majorana particle which is defined as is
its own anti particle and is characterized by only two independent
particle states of the same mass ($\nu^{~}_{\rm L}$ and
$\bar{\nu}^{~}_{\rm R}$ or $\nu^{~}_{\rm R}$ and
$\bar{\nu}^{~}_{\rm L}$). A Majorana mass term which violates both
the law of total lepton number conservation and that of individual
lepton flavor conservation can be written either as
\begin{equation}
 \frac{1}{2} \overline{\nu}_{a_{L}} M_L {\nu}^c_{a_R}
+ h.c.,
\end{equation}
or as
\begin{equation}
\frac{1}{2} \overline{\nu}^c_{a_L} M_R {\nu}_{a_R} + h.c.,
\end{equation}
where $M_l$ and $M_R$ are complex symmetric matrices leading to
the famous see-saw mechanism \cite{seesaw}-\cite{seesaw5}, given
by \be M_{\nu}=-M_{\nu D}^T\,(M_R)^{-1}\,M_{\nu D},
\label{seesaweq} \ee \noindent where $M_{\nu D}$ and $ M_R$ are
respectively the Dirac neutrino mass matrix and the right-handed
Majorana neutrino mass matrix. This mechanism requires the
inclusion of right-handed neutrinos with very large Majorana
masses, therefore inducing a very small mass for the left-handed
neutrinos. Thus, the generation of masses in neutrinos is not
straight-forward as they may have either the Dirac masses or the
more general Dirac-Majorana masses. Further, when discussing
texture possibilities textures are imposed on $M_{\nu D}$, unlike
many other attempts \cite{flbaatt1}-\cite{flbaatt8} in the
literature where texture is imposed on $M_{\nu}$.

In the case of the Majorana neutrinos, there are extra phases
which cannot be removed, therefore, the above matrix $V_{PMNS}$
takes the following form \beqn {\left( \ba{ccc} c_{12} c_{13} &
s_{12} c_{13} & s_{13}e^{-i {\delta}_l} \\ - s_{12} c_{23} -
c_{12} s_{23} s_{13} e^{i {\delta}_l} & c_{12} c_{23} - s_{12}
s_{23} s_{13} e^{i {\delta}_l} & s_{23} c_{13}
\\ s_{12} s_{23} - c_{12} c_{23} s_{13} e^{i \delta_l} & - c_{12}
s_{23} - s_{12} c_{23} s_{13} e^{i { \delta}_l} & c_{23} c_{13}
\ea \right)} \left( \ba{ccc} e^{i \alpha_1/2} & 0 & 0 \\ 0 &e^{i
\alpha_2/2} & 0 \\ 0 & 0  & 1 \ea \right),\nonumber \\ \eeqn
 where $\delta_{l}$ is the
Dirac-like CP violating phase in the leptonic sector and
$\alpha_1$ and $\alpha_2$ are the Majorana phases which do not
play any role in neutrino oscillations.

{\section{Experimental status of neutrino masses and mixing
parameters} \label{input}} While carrying out an analysis
regarding exploring the compatibility of neutrino mass matrices
with the recent data, one needs to keep in mind the experimental
constraints imposed by the relationship between mass matrices and
their corresponding mixing matrices. To facilitate our discussion
in this regard, we present the status of relevant data in the
lepton sector. The $ 3 \sigma$ confidence level ranges of the
neutrino oscillation parameters obtained in a latest global three
neutrino oscillation analysis carried out by Fogli {\it et al}.
\cite{fogli2012} have been presented in Table (\ref{data}).

\begin{table}[ht]
\centering
\begin{tabular}{|c| c|}
\hline Parameter  & $3 \sigma$ range  \\ [0.5ex] \hline

$\Delta m_{sol}^2$ $[ 10^{-5} eV^2]$  & (6.99-8.18) \\ \hline
$\Delta m_{atm}^2$ $[ 10^{-3} eV^2]$ & (2.19-2.62)(NH);
(2.17-2.61)(IH)  \\ \hline $sin^2 \theta_{13}$ $[10^{-2}]$  &
(1.69-3.13)(NH); (1.71-3.15) (IH)  \\ \hline $sin^2 \theta_{12}$
$[10^{-1}]$  & (2.59-3.59)  \\ \hline $sin^2 \theta_{23}$
$[10^{-1}]$  & (3.31-6.37)(NH);(3.35-6.63)(IH)  \\ \hline
\end{tabular}
\caption{Current data for neutrino mixing parameters from global
fits \cite{fogli2012}.} \label{data}
\end{table}

While carrying out the analysis, the magnitudes of atmospheric and
solar neutrino mass square differences, defined as $m_2^2-m_1^2$
and $m_3^2-\frac{(m_1^2+m_2^2)}{2}$ respectively, are allowed full
variation within their $3\sigma$ ranges. The lightest neutrino
mass, $m_1$ for the case of normal hierarchy (NH) and $m_3$ for
the case of inverted hierarchy (IH), is considered as the free
parameter while the other two masses are obtained using the
following relations, \be NH:~~m_2^2=\Delta m_{sol}^2 +m_1^2,~~
m_3^2=\Delta m_{atm}^2 + \frac{(m_1^2+m_2^2)}{2}, \ee \be
IH:~~m_2^2=\frac{2 (m_3^2+\Delta m_{atm}^2)+ \Delta
m_{sol}^2}{2},~~ m_1^2=\frac{2 (m_3^2+\Delta m_{atm}^2)-\Delta
m_{sol}^2}{2}.\ee

It should be noted that while carrying out analyses of different
texture specific mass matrices, we have also imposed the condition
of `naturalness' \cite{minireview} so as to keep the quark-lepton
similarity in this regards. Further, the phases
$\phi_1=\alpha_{\nu D}-\alpha_l$, $\phi_2= \beta_{\nu D}-\beta_l$
and the elements $D_{l, \nu}$, $C_{l, \nu}$ are considered to be
free parameters. In the absence of any constraint on the phases,
$\phi_1$ and $\phi_2$ have been given full variation from 0 to
$2\pi$. Although $D_{l, \nu}$ and $C_{l, \nu}$ are free
parameters, however, they have been constrained such that
diagonalizing transformations $O_l$ and $O_{\nu}$ always remain
real.

Before presenting the results, we would like to mention that
unlike the quark case, wherein it has been shown that texture 4
zero  Fritzsch like matrices are perhaps the only compatible
matrices with data \cite{minireview,singreview},
\cite{pramana}-\cite{prd}, in the case of leptons, we cannot
arrive at this kind of conclusion. In the sequel, we present an
overview of the viability of different textures for Dirac as well
as Majorana nature of neutrinos.

\section{Viable texture specific lepton mass matrices}In the  context of quarks
it is well known that texture 6 zero mass matrices are completely
ruled out by the existing data. Interestingly, in case we consider
Dirac like neutrinos, texture 6 zero or minimal texture is also
ruled out for normal/ inverted hierarchy and degenerate scenario
of neutrino masses. However, for Majorana neutrinos inverted
hierarchy and degenerate scenario are ruled out whereas in the
case of normal hierarchy, there are several compatible
combinations with the current neutrino oscillation data. For a
detailed discussion, we refer the readers to \cite{singreview}.

Coming to the cases of non-minimal textures, i.e., the texture 5
zero and texture 4 zero mass matrices, we present our conclusions
from our recent analyses \cite{arxive1}, \cite{arxive2}. To begin
with, we first discuss the texture 5 zero and texture 4 zero
lepton mass matrices for the case of Dirac neutrinos.
Corresponding to this, a detailed and comprehensive analysis has
been carried out for normal/ inverted hierarchy and degenerate
scenario of neutrino masses. In this context, for texture 5 zero
mass matrices, the analysis has been carried out for $D_l=0$,
$D_{\nu}\neq 0$ as well as $D_l\neq 0$, $D_{\nu}= 0$ cases,
corresponding to all the viable classes. For class I, mentioned in
Table I, inverted hierarchy is ruled out for both the cases,
whereas normal hierarchy is viable for the $D_l=0$, $D_{\nu}\neq
0$ case. For class II, normal hierarchy is viable for both the
cases while the inverted hierarchy is ruled out for the case
$D_l=0$, $D_{\nu}\neq 0$. Finally, for class III we find that
inverted hierarchy is viable for the case $D_l=0$, $D_{\nu}\neq
0$, while the normal hierarchy is compatible with the $D_l \neq
0$, $D_{\nu}= 0$  case. It may be mentioned that Class IV is not
phenomenologically viable due to de-coupling of one of the
generations.

Coming to the texture 4 zero case, due to the availability of an
additional parameter large number of viable possibilities emerge.
Without getting into the details of these possibilities, we would
like to mention only broad conclusions in this regard.
Interestingly, unlike the case of texture 6 zero mass matrices,
both inverted hierarchy and degenerate scenario are not ruled out
for all the classes of texture specific mass matrices mentioned in
Table 1. For the case of normal hierarchy, it seems mass matrices
corresponding to all the classes are compatible with the data.
However, inverted hierarchy is ruled out for Class I but
compatible with Class II and Class III. Similarly, degenerate
scenario of neutrino masses is compatible only with Class III.

Coming to the case of texture 5 zero and texture 4 zero mass
matrices for neutrinos being Majorana particles. To begin with, we
consider texture 5 zero lepton mass matrices, for both the cases,
viz. $D_l=0$, $D_{\nu}\neq 0$ as well as $D_l\neq 0$, $D_{\nu}= 0$
for matrices mentioned in Class II and Class III of Table 1. For
class II, normal hierarchy is viable for both the cases, while the
inverted hierarchy seems to be ruled out for the case, $D_l=0$,
$D_{\nu}\neq 0$. Finally, for texture 5 zero mass matrices
pertaining to class III, we find that inverted hierarchy is viable
for the case $D_l\neq 0$, $D_{\nu}= 0$, while the normal hierarchy
is compatible with the $D_l=0$, $D_{\nu}\neq 0$ case.

It may be mentioned that the number of viable possibilities is
understandably  quite large.  The analysis reveals that the
Fritzsch like texture two zero lepton mass matrices are compatible
with the recent lepton mixing data pertaining to normal as well as
inverted neutrino mass hierarchies. Interestingly, one finds that
both the normal as well as inverted neutrino mass hierarchies are
compatible with texture four zero mass matrices pertaining to
class II and III  of Table 1 contrary to the case for texture four
zero mass matrices pertaining to class I wherein inverted
hierarchy seems to be ruled out. Interestingly for classes I and
II, the degenerate neutrino mass scenario is incompatible, whereas
it is compatible for mass matrices in Class III.

It is interesting to add that in the context of quarks, it has
been recently shown \cite{prd} that texture 4 zero Fritzsch like
mass matrices and their permutations, compatible with the Weak
Basis transformations, provides a unique texture in agreement with
the data. In case of leptons also, we have seen that texture 4
zero matrices are compatible with data for Dirac as well as
Majorana neutrinos. Therefore, we can conclude that Fritzsch like
texture 4 zero matrices may provide vital clues for the
fundamental theories of flavor physics.

\section{Summary and conclusions}
A broad based survey of the texture specific lepton mass matrices
has been presented. It seems that in the case of Dirac neutrinos,
texture 6 zero mass matrices are ruled out. However, this is not
true in the case of Majorana neutrinos. Lesser than texture 6
zeros, we find compatibility of the mass matrices with data for
both the kind of neutrinos and for all kind of neutrino mass
hierarchies.

\vskip 0.5cm {\bf Acknowledgements} \\ G.A. would like to
acknowledge DST, Government of India (Grant No:
SR/FTP/PS-017/2012) for financial support.  M.G. and P.F.  would
like to acknowledge CSIR, Govt. of India,(Grant
No:03:(1313)14/EMR-II) for financial support. S.S. acknowledges
the Principal, GGDSD College, Sector 32, Chandigarh.   P.F. and
S.S. acknowledge the Chairperson, Department of Physics, P.U., for
providing facilities to work.

\end{document}